\documentclass[a4paper, 11pt]{llncs}
\usepackage{amsmath,amssymb}
\usepackage{url}
\usepackage{graphicx} 
\usepackage{epsfig}

\newcommand{\prob}{\mbox{\bf Pr}}
\newcommand{\ex}{\mbox{\bf E}}

\begin{document}
%
\title{{\bf A probabilistic key agreement scheme for sensor networks without key
predistribution \thanks{Partially supported by the IST Programme
of the European Union under contact number IST-2005-15964 (AEOLUS)
and by the ICT Programme of the European Union under contract
number ICT-2008-215270 (\textsf{FRONTS}).}}}
%
%
%
\author{
V. Liagkou \inst{1,3} \and E. Makri \inst{4} \and P. Spirakis
\inst{1,3} \and Y.C. Stamatiou \inst{2,3}}
%
%

%
\institute{University of Patras, Dept. of computer Engineering,
26500, Rio, Patras, Greece \and Mathematics Department, 451 10,
Ioannina, Greece. e-mail: istamat@uoi.gr \and Research and
Academic Computer Technology Institute, N. Kazantzaki, University
of Patras, 26500, Rio, Patras, Greece \and University of the
Aegean, Department of Mathematics, 83000, Karlovassi, Samos,
Greece. }

\pagestyle{plain}

\maketitle



\begin{abstract}

The dynamic establishment of shared information (e.g. secret key)
between two entities is particularly important in networks with no
pre-determined structure such as wireless sensor networks (and in
general wireless mobile ad-hoc networks). In such networks, nodes
establish and terminate communication sessions dynamically with
other nodes which may have never been encountered before, in order
to somehow exchange information which will enable them to
subsequently communicate in a secure manner.
In this paper we give and theoretically analyze a series of
protocols that enables two entities that have never encountered
each other before to establish a shared piece of information for
use as a key in setting up a secure communication session with the
aid of a shared key encryption algorithm. These protocols do not
require previous pre-distribution of candidate keys or some other
piece of information of specialized form except a small seed
value, from which the two entities can produce arbitrarily long
strings with many similarities.

\end{abstract}


\section{Introduction}

Wireless Sensor Networks (WSNs) have some constraints, with regard
to battery life, processing, memory and
commnication~(\cite{key-6}) capacity, and as such are deemed
unsuitable for public crypto-based systems. Thus, symmetric key
cryptosystems are more appropriate for these types of networks,
but lead to problems with key distribution. These problems are
mitigated with key {\em pre-distribution schemes}, in which
candidate keys are distributed to members of the network before
the start communication.

Many innovative and intuitive key pre-distribution schemes for
WSNs have been proposed for solving the problem of key
distribution in sensor networks. On the two ends of the spectrum
are key pre-distribution schemes that use a single master key as
the encryption key distributed amongst all the nodes, and all
pairwise keys, where a unique key exists for every pair of
sensors. The former provides the most efficient usage of memory
and scales well, but an attack on one node compromises the whole
network, whereas the latter provides excellent resilience but does
not scale well. In addition, schemes exist which are in essence
probabilistic, relying on the fact that any two neighbouring nodes
have some probability $p$ of successfully completing key
establishment. Some such schemes are presented in the sequel, but
the list is by no means exhaustive.
The authors of~\cite{key-1} propose a key distribution scheme
which consists of three phases. In the first phase, namely the
\emph{key pre-distribution phase}, $k$ random keys are drawn from
a generated pool of $P$ keys, and are preloaded onto the sensor
nodes before their deployment. These $k$ keys constitute the key
ring of each sensor node. The key identifiers of the key rings are
loaded onto controller nodes, along with the identifiers of the
corresponding nodes and the keys shared between the controller
nodes and the sensor nodes. Once the nodes are deployed, the
second phase takes place, the \emph{shared-key discovery phase},
which allows nodes to discover their neighbours within
communication range and with which they share keys. This phase
establishes the topology of the sensor network, as seen by the
routing layer. The third, and final phase, is the \emph{path-key
establishment} phase which assigns a path-key to pairs of sensor
nodes in wireless communication range that do not share a common
key.
The authors of~\cite{key-2} present an alternative to this scheme, the
\emph{q-composite random key predistribution scheme}, whereby
instead of one single common key to be shared between two nodes,
$q$ keys are required, resulting in strengthened securty under
small scale attacks, trading off increased vulnerability on large
scale attacks. The authors also propose the establishment of the
key-path along multiple paths, and a random-pairwaise keys scheme
which provides, amongst others, node-to-node authentication.
The authors of~\cite{key-3} also take the scheme of~\cite{key-1}
and modify it by taking into consideration that different
locations of nodes require different security needs, and as such
propose a subgrouping approach in order to isolate any node
captures that might take place in specific subgroups. The scheme
also provides scalability for key pre-distribution, by taking into
account the probability of node compromise for each of the
subgroups.
Chan and Perrig~(\cite{key-4}) address the lack of scalability of
existing symmetric-key key distribution protocols, and increase
the security against node compromise. In their proposed
\emph{PIKE} protoccol, sensor nodes are used as trusted
intermediaries to establish shared keys between nodes in the
wireless sensor network. The authors of~\cite{key-9} propose a
pair-wise key pre-distribution scheme, which is based on the
scheme proposed by~\cite{key-17}. Their work involves the the
pre-loading of crypto shares from multiple key spaces onto each
sensor node, and after deployment, two nodes may establish a
shared key between them only if they contain crypto shares from
the same key space, leading to an improvement in scalability.
Finally, the same authors~(\cite{key-10}) propose a scheme which
utilises deployment knowledge, and generates pairwise keys between
nodes and their neighbours, guaranteeing that each node can
establish a secure communications link with its neighbours post
deployment, with a high probability.
The previous solutions all pre-suppose that the sensor nodes have
been loaded with some pre-existing information (i.e. the key, or
sets of keys) prior to network deployment, except for Liu and
Cheng~(\cite{key-8}). They propose a self-configured scheme
whereby no prior knowledge is loaded onto the sensor nodes, but
shared keys are computed amongst the neighbours. The authors
propose \emph{SBK} which is also a topology adaptive scheme,
achieves high connectivity with a small storage overhead, and
\emph{iSBK}, and improved version of \emph{SBK}.

In this paper, we propose a key agreement scheme whereby network
nodes are not pre-loaded with candidate keys, but generate pairs
of symmetric keys from two, initially, random bits strings. The
initial research conducted~(\cite{key-5}) proposed a protocol that
involved the examination of random positions of subsets of size
$k$, and the elimination of a random position if the two bit
strings were found to disagree on more than half the examined
positions. In that paper, however, the nodes cannot secretly
compute the number of differing positions, a problem that is
resolved in the present paper using secret circuit computations.
In addition, the present protocols do not eliminate differing bits
but flips them, depending on the number of bit difference in the
examined subset of $k$ bits. This leads to a different stochastic
process that called for a different theoretical analysis. The
scheme we propose has the following four properties: (i)
scalability, since new nodes may enter the network whenever they
desire without the need to equip them with some suitable key set
(ii) connectivity, since any two nodes of the network can reach a
large degree of similarities on the strings the possess (iii) no
storage overhead, since the string sizes are independent on the
network size and (iv) resilience against node captures, since
capturing any number of nodes of the network does not affect the
strings other nodes possess. On the negative side, the protocol we
give have some increased communication overhead. However, once two
nodes reach a position where their two strings have sufficiently
much similarity, they can create from them a large number of
candidate keys (possibly using error correction techniques to
create even more similarities) for their current and future
communication needs.


\section{The bit-similarity problem}
\label{protocol}

Two entities, say 0 and 1, initially possess an $N$-bit string,
$X_N^0$ and $X_N^1$ respectively. The entities' goal is to
cooperatively transform their strings so as to increase the
percentage of positions at which their strings contain the same
bits, which we denote by $X(i)$, with $i$ being the time step of
the protocol they execute. Then $X(0)$ is the initial percentage
of the positions at which the two strings are the same.

Below we provide a randomized protocol in which the two entities
examine randomly chosen subsets of their strings in order to see
whether they differ in at least half of the places. If they do,
one of the entities (in turn) randomly flips a subset of these
positions. This process continues up to a certain, predetermined
number of steps. The intuition behind this protocol is that when
two random substrings of two strings differ in at least half of
their positions, then flipping some bits at random in one of the
substrings is more likely to increase the percentage of
similarities between the strings than to decrease it. In the
description of the protocol $X_N^c[S]$ denotes a substring of
string $X_N^c$ defined by the position set $S$.
\medskip
\begin{small}
\noindent Protocol for user $U_{c}, c = 0, 1$

\noindent Protocol parameters known to both communicating parties:
(i) $k,l$, the subset sizes, (ii) $T$, the number of protocol
execution steps, (iii) the index (bit position) set $N$, (iv) The
circuit $C$ with which the two entities jointly compute whether
there are at least $\lceil k/2 \rceil$ similarities between
randomly chosen subsets of their strings.
\begin{enumerate}
\item $i \leftarrow 1$ /* The step counter. */
\item {\bf while $i \leq T$} /* $T$ is a predetermined time step limit (discussed in Section~\ref{efficiency}).*/
\item \hspace{0.15cm} {\bf begin} /* while */
\item \hspace{0.35cm} $S \leftarrow \mbox{JOINT\_RAND}(k, \{1, \ldots,N \})$ /* Shared random set of $k$ positions. See text. */
\item \hspace{0.35cm} same\_pos $\leftarrow C(X_N^c[S],X_N^{(c + 1 \mod
2)}[S])$ /* A secret computation of number of positions with same
contents (see Section~\ref{sec_comp}). */
\item \hspace{0.35cm} {\bf if} (same\_pos $\geq \lceil \frac{k}{2} \rceil$ {\bf and}  $\mbox{odd}(i + c)$) {\bf then} /* Users 0 and 1 alternate. */
\item \hspace{0.55cm} {\bf begin}
\item \hspace{0.7cm} $SF \leftarrow \mbox{RAND}(l, S)$ /* Random set of $l$ positions from within $S$ to be flipped by the user whose turn it is to flip. */
\item \hspace{0.7cm} flip the bits of $X_N^c[SF]$
\item \hspace{0.55cm} {\bf end}
\item \hspace{0.35cm} SYNCHRONIZE /* Users 0 and 1 wait to reach this point simultaneously (barrier synchronization). */
\item \hspace{0.35cm} $i \leftarrow i + 1$
\item \hspace{0.15cm} {\bf end} /* while */
\end{enumerate}
\end{small}
An important requirement of the protocol is the existence of a
random number generator at each party which produces the same
values, started from a (small) seed value shared by both parties.
This is in order to allow the two parties to avoid sending over
the communication channel the chosen substring positions, as they
are generated in synchrony by the two parties (see Line 4 of the
protocol).

With regard to the protocol, we are interested in the evolution of
the random variable $X(i)$, i.e. the percentage of the positions
of $X_N^0$ and $X_N^1$ at which they are the same, after the $i$th
step of the protocol.

\section{Secret two-party function computation}
\label{sec_comp}

During the execution of the protocol, it is necessary for the two
communicating parties to see whether they agree on at least half
of the positions they have chosen to compare (line 13-14 of the
protocol). Thus, the two parties need to perform a computation:
compute the number of positions on which the corresponding bits in
the two chosen subsets of k bits are the same. This is an instance
of an important, general problem in cryptography: {\em Secure
Computation}.

Secure computation techniques allow two (or more) communicating
parties to compute a function on their inputs so that nothing is
revealed to each party except what can be inferred from own input
and the computed value. Moreover, no information is revealed about
the bit values to a party not involved in the computation (e.g.
eavesdropper).

More formally, let $A$ and $B$ be two parties with inputs of $n_A$
and $n_B$ bits respectively. The objective is to jointly compute a
function $f: \{0,1\}^{n_A} \times \{0,1\}^{n_B} \rightarrow
\{0,1\}$ on their inputs. The issue, here, is that A and B cannot,
simply, exchange their inputs and compute the function since they
will learn each other's inputs, something that is not desirable in
a secure computation setting. More importantly, even it A and B
are willing to share their inputs, they would not allow an
eavesdropper to acquire these inputs too. This leads to the
problem of {\em secure function computation}.

In our context, we consider the following two Boolean functions:
\begin{eqnarray}
& & f_{r}: \{0,1\}^{k} \times \{0,1\}^{k} \rightarrow \{0,1\}
\mbox{ with } w_A, w_B \in \{0,1\}^{k} \mbox{ and } 0 \leq r \leq
k : \nonumber
\\ & & f(w_A, w_B) = \lbrace{\begin{array}{cc} 1, & \mbox{ if }
X(w_A,w_B) \geq  r  \\ 0, & \mbox{ otherwise }
\end{array}}
\end{eqnarray}
We are interested in $r = \lceil \frac{k}{2} \rceil$.
\begin{eqnarray}
& & f_{X}: \{0,1\}^{k} \times \{0,1\}^{k} \rightarrow
\{0,1\}^{\lceil \log_2(k) \rceil} \mbox{ with } w_A, w_B \in
\{0,1\}^{k} : \nonumber
\\ & & f_X(w_A, w_B) = x, \mbox{ with } x = X(w_A,w_B) \mbox{ written in binary}.
\end{eqnarray}
The function $f_X$ is, strictly, an ordered tuple $(f_{X}^0,
f_{X}^{1}, \ldots, f_{X}^{{{\lceil \log_2(k) \rceil - 1}}})$ of
$\lceil \log_2(k) \rceil$ 1-bit Boolean functions, where the
function $f_X^{i}$ computes the $i$th most significant bit of
$x=X(w_A,w_B)$ (with $i$ = 0 we take the most significant bit and
with $i = \lceil \log_2(k) \rceil - 1$ we take the least
significant bit).
Using techniques from oblivious function computation
(see~\cite{Cramer99} for a survey on these techniques), we can
prove that the computation of $f_r$ and $f_X$ can be done with
randomized protocols using $O(|C_{f_r}|)$ and $O(|C_{f_{X}}|)$
communication steps respectively, with $C_{f_r}$ and $C_{f_X}$
being the Boolean circuits that are employed for the computation
of $f$ and $f_X$ respectively. Since both $f_r$ and $f_X$ are
easily seen to be polynomial time computable Boolean functions, we
can construct for their computations circuits of size polynomial
in their input sizes, i.e. circuits $C_{f_r}$ and $C_{f_X}$ such
that $|C_{f_r}| = O(k^{c_1})$ and $|C_{f_{X}}| = O(k^{c_2})$, with
constants $c_1, c_2 \geq 0$. Since $k$ is considered a fixed
constant, we conclude that we can compute $f_k$ and $f_{X}$ in a
constant number of rounds. The number of random bits needed by
each step of the randomized protocol is in both cases $O(k)$ and,
thus, constant.

To sum up, the functions $f_r$ and $f_X$ can, both, be evaluated
on two $k$-bit inputs $w_A, w_B$ held by two parties $A,B$ using a
constant number of rounds and a constant number of uniformly
random bits. In what follows, we will assume that the
communicating parties use the function $f_{\lceil
\frac{k}{2}\rceil}$.

With regard to the required randomness, we assume that each of the
two parties has a true randomness source, i.e. a source of
uniformly random bits. Such a randomness source can be easily
built into modern devices.
This randomness source is necessary in order to implement the
randomized oblivious computation protocols for the computation of
the function $f_{\lceil \frac{k}{2} \rceil}$. In addition, it will
be used in order to produce the randomly chosen positions, are
required by Step 6 of the protocol. Since each position can range
from 1 up to $N$ (the string size), to form a position index we
need to draw $\lceil \log_2(N)\rceil$ random bits. Alternatively,
if we allow the two parties to share a small (in relation to $N$)
seed, they can produce the random positions in synchronization
and, thus, avoid sending them over the communication channel.

\section{Theoretical analysis of the protocol}
\label{analysis}

In order to track the density of positions where two strings
agree, we will make use of Wormald's theorem (see~\cite{Wor99}) to
model the probabilistic evolution of the protocol described in
Section~\ref{protocol} using a deterministic function which stays
provably close to the real evolution of the algorithm. The
statement of the theorem is in the Appendix for completeness. What
the theorem essentially states is that if we are confronted with a
number of (possibly) interrelated random variables (associated
with some random process) such that they satisfy a Lipschitz
condition and their expected fluctuation at each time step is
known, then the value of these variables can be approximated using
the solution of a system of differential equations. Furthermore,
the system of differential equations results directly from the
expressions for the expected fluctuation of the random variables
describing the random process.

We will first prove a general lemma that gives the probability of
increasing the similarity between two strings through flipping, at
random, the contents of a certain number of positions.

\begin{lemma}
Let $w_1,w_2$ be two strings of 0s and 1s of length $k$. Let also
$j$, $0 \leq j \leq k$, be the number of places in which the two
strings differ. Then if $l$ positions of one string are randomly
flipped, the probability that $s$ of them are differing positions
is the following:
\begin{equation}
P_{k,j,l,s}= \frac{{j \choose s} {k-j \choose l-s}}{{k \choose
l}}. \label{eqflips}
\end{equation}

\label{problemma}

\end{lemma}

\noindent{\bf Proof} In Equation~(\ref{eqflips}) the denominator
is the number of all subsets of positions of cardinality $l$ of
the $k$ string positions while the numerator is equal to the
number of partitions of the $l$ chosen positions such that $s$ of
them fall into the $j$ differing positions and the remaining $l-s$
fall into the remaining $k-j$ non-differing positions of the two
strings. Thus their ratio gives the desired probability. \hfill
$\Box$

The following lemma, which is easy to prove based on general
properties of the binomial coefficients, provides a closed form
expression for a sum that will appear later in some probability
computations.

\begin{lemma}
The following identity holds:
\begin{equation}
\sum_{s=0}^{l}(2s-l)\frac{{j \choose s} {k-j \choose l-s}}{{k
\choose l}} = \left(\frac{2j}{k}-1\right)l.
\end{equation}

\label{sumlemma}
\end{lemma}
We will now derive the deterministic differential equation that
governs the evolution of the random variable $X(i)$ manipulated by
the protocol in Section~\ref{protocol} using Wormald's theorem.

\begin{theorem}
The differential equation that results from the application of
Wormald's theorem on the quantity $X(i)$ (places of agreement at
protocol step $i$) as it evolves in the agreement protocol is the
following:

\begin{eqnarray}
& & \ex[X(i+1)-X(i)] = \nonumber \\ && \sum_{j=\lceil \frac{k}{2}
\rceil }^k \sum_{s=0}^{l} [(s-(l-s)) P_{k,j,l,s}]P_{n,n-X(i),k,j}.
\label{diffeq}
\end{eqnarray}

\label{general}
\end{theorem}

\noindent {\bf Proof} We will determine the possible values of the
difference $X(i+1)-X(i)$ along with the probability of occurrence
for each of them.

The protocol described in Section~\ref{protocol} flips $l$
positions within the $k$ examined positions, whenever these $k$
positions contain $j \geq \lceil \frac{k}{2} \rceil$ differing
positions in the two strings. From the flipped $l$ positions, is
$s$ of them ($0 \leq s \leq l$) are disagreement positions, then
the two strings will have gained $s$ agreement positions, losing
$l-s$. The net total is $s-(l-s)$. The probability that this total
occurs, for a specific value of $s$ and a specific value of $j$ is
equal to $P_{k,j,l,s} P_{n,n-X(i),k,j}$. Summing up over all
possible values of $s,j$ we obtain~(\ref{diffeq}). \hfill $\Box$

\begin{corollary}
The following holds:

\begin{eqnarray}
& & \ex[X(i+1)-X(i)] \nonumber \\ &&= \sum_{j=\lceil \frac{k}{2}
\rceil }^k l \left(\frac{2j}{k} - 1 \right)\frac{{n - X(i) \choose
j}{X(i) \choose k-j }}{{n \choose k}}. \label{diffeq2}
\end{eqnarray}
\label{diffeqcor}
\end{corollary}

\noindent{\bf Proof} Equation~(\ref{diffeq2}) follows
from~(\ref{diffeq}) using Equation~(\ref{eqflips}) of
Lemma~\ref{problemma} with $k=n,l=k,s=j,n-j=X(i)$, in conjunction
with Lemma~\ref{sumlemma}.

\begin{corollary}
Using Wormald's Theorem (Theorem~\ref{wormald}), the evolution of
the random variable $X(i)$ whose mean fluctuation is given
in~(\ref{diffeq2}) can be approximated by the following
differential equation:
\begin{eqnarray}
& & \frac{dx(t)}{dt} \nonumber \\ &&= \sum_{j=\lceil \frac{k}{2}
\rceil }^k l \left(\frac{2j}{k} - 1 \right) {k \choose j} [1-
x(t)]^{j} x(t)^{k-j}.\label{diffeqapp}
\end{eqnarray}
\end{corollary}
\noindent{\bf Proof} By applying the approximation $$ {N \choose
k} = \frac{N^k}{k!}\left(1 + O\left(\frac{1}{N}\right)\right) $$
of the binomial coefficients which is valid for $k=O(1)$ on the
three binomials which appear on the the right-hand side
of~(\ref{diffeq2}), we obtain the following:
$$ \frac{{n - X(i) \choose j}{X(i) \choose k-j }}{{n \choose k}}
\simeq {k \choose j} \left(1 - \frac{X(i)}{n} \right)^j
\left(\frac{X(i)}{n}\right)^{k-j}.$$ Using Wormald's theorem, we
make the correspondence $x(t) = \frac{X(i)}{n}$ and
$\frac{dx(t)}{dt} = \ex[X(i+1)-X(i)]$, which results in the
required differential equation~(\ref{diffeqapp}). \hfill $\Box$

\section{Efficiency of the protocol}
\label{efficiency}

From Equation~(\ref{diffeqapp}) we see that the percentage of
similar positions, represented by the function $x(t)$, is a
monotone increasing function since its first derivative is always
positive. In what follows, we will estimate how fast this
percentage increases depending on its initial value $x(0)$ as well
as the parameters $l$ and $k$.

\begin{lemma}
The solution $x(t)$ to the differential equation given
in~(\ref{diffeqapp}) is monotone increasing.
\end{lemma}

\noindent {\bf Proof} From the differential equation, we see that
the first derivative of the function $x(t)$, which is equal to the
right-hand side of the differential equation, is strictly
positive, since $0 < x(t) < 1$. Thus, the function $x(t)$ is
monotone increasing. \hfill $\Box$

\begin{lemma}
Let $x(t_1)$ be the value of the function $x(t)$ at time instance
$t_1$ and $x(t_2)$ be the value at time instance $t_2$, $t_1 <
t_2$. Let, also, $c(t_1)$ be the absolute value of the point at
which the tangent line to the point $(t_1,x(t_1))$ of $x(t_1)$
cuts the $t$-axis and $c(t_2)$ the corresponding value for $t_2$.
Let, also,
\begin{equation}
p(x)=\sum_{j=\lceil \frac{k}{2} \rceil }^k l \left(\frac{2j}{k} -
1 \right) {k \choose j} (1 - x)^{j} x^{k-j}. \label{pol}
\end{equation}
Then
\begin{equation}
p(x(t_1)) = \frac{x(t_1)}{c(t_1)+t_1}, p(x(t_2)) =
\frac{x(t_2)}{c(t_2) + t_2}. \label{tan}
\end{equation}

\label{tanlemma}
\end{lemma}

\begin{figure}[ht]
    \centering
        \includegraphics[width=0.7\textwidth]{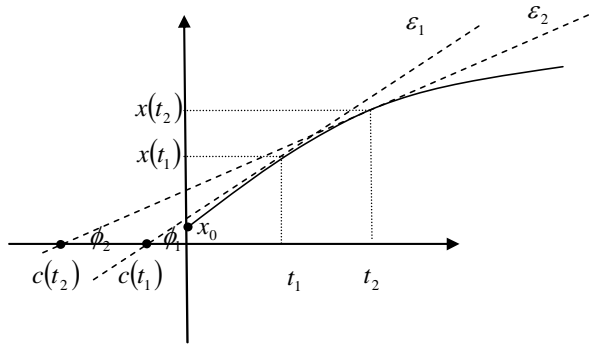}
    \caption{The two tangent lines for the proof of the Theorem}
    \label{tang}
\end{figure}

\noindent {\bf Proof}
Let $\epsilon_1$ and $\epsilon_2$ be the two tangent lines to the
function $x(t)$ at the points $(t_1, x(t_1))$ and $(t_2, x(t_2))$
respectively, as shown in Figure~\ref{tang}. Due to the
monotonicity of $x(t)$, the points at which the two lines
intersect with the $t$-axis are negative. Let $c(t_1)$ and
$c(t_2)$ be the absolute values of these two points for lines
$\epsilon_1$ and $\epsilon_2$ respectively.
Then from the two right angle triangles that are formed we have
$\tan(\phi_1) = \frac{x(t_1)}{c(t_1)+t_1}$ and $\tan(\phi_2) =
\frac{x(t_2)}{c(t_2)+t_2}$. From the definition of the derivative,
$\tan(\phi_1) = \frac{dx(t)}{dt}|_{t_1} $ and $\tan(\phi_2) =
\frac{dx(t)}{dt}|_{t_2}$. From~(\ref{diffeqapp}) and~(\ref{pol}),
we have $\frac{dx(t)}{dt} = p(x(t))$ and, thus, the statement of
the lemma follows. \hfill $\Box$

\begin{theorem}
Let $t'$ be the time instance at which $x(t') = h x(0)$, with $1
\leq h \leq \frac{1}{x(0)}$. Then
\begin{equation}
t_2 \leq \frac{x(0)}{p(hx(0))} \cdot(h - 1). \label{up}
\end{equation}
\end{theorem}

\noindent {\bf Proof} We set $t_1 =0, t_2 = t'$ and $x(t') = h
x(0)$ in Lemma~\ref{tanlemma} and we obtain, from
Equations~(\ref{tan}) the following:
\begin{eqnarray*}
p(x(0)) = \frac{x(0)}{c(0)}, p(h x(0)) = \frac{h x(0)}{c(t') +
t'}.
\end{eqnarray*}
From these equations we obtain the following:
$$ \frac{p(x(0))}{p(h x(0))} = \frac{c(t')+t'}{h c(0)}. $$
Solving for $t'$ we obtain the following:
\begin{equation}
t' = \frac{h p(x(0)) c(0)}{p(h x(0))} - c(t'). \label{prime}
\end{equation}
Since $p(x(0))$ is the inclination of the tangent line to $x(t)$
at the point $(0,x(0))$, it holds that $p(x(0)) =
\frac{x(0)}{c(0)}$. Thus,~(\ref{prime}) becomes
\begin{equation}
t' = \frac{h x(0)}{p(h x(0))} - c(t'). \label{prime2}
\end{equation}
Since $x(t)$ is monotone increasing, the point at which the
tangent to this function cuts the $x(t)$-axis at any point is
greater than or equal to $x(0)$ (see Figure~\ref{tang}). Let $x_c$
be this point. Then $p(x(t')) = \frac{x_c}{c(t')}$ or, since
$x(t') = h x(0)$, $p(h x(0)) = \frac{x_c}{c(t')}$. Since $x_c \geq
x(0)$, $c(t') \geq \frac{x(0)}{p(h x(0))}$. Thus,
from~(\ref{prime2}), we obtain $$ t' \leq \frac{h x(0)}{p(h x(0))}
- \frac{x(0)}{p(h x(0))} = \frac{x(0)}{p(h x(0))} \cdot (h - 1)$$
which is the required. \hfill $\Box$

\begin{lemma}
The following lower bounds hold for the polynomial in~(\ref{pol}):

\noindent If $x < 1/2$ then
\begin{eqnarray}
p(x) \geq  l[ x^k + 1 - 2x].
\end{eqnarray}

\noindent If $x \geq 1/2$ then
\begin{eqnarray}
p(x) \geq \frac{l}{k} (1-x)^k {k \choose \lceil \frac{k}{2}
\rceil} \lceil \frac{k}{2} \rceil.
\end{eqnarray}

\end{lemma}

\noindent {\bf Proof} In~(\ref{pol}), allowing the sum index to
cover all the range from 1 to $k$ reduces the value of the sum
since it adds negative terms. Thus
$$ p(x) \geq l \sum_{j=1}^{k} \left (\frac{2j}{k}-1\right) {k
\choose j} (1-x)^j x^{k-j}. $$ Since the sum evaluates to $l[ x^k
+ 1 - 2x]$ the first statement of the lemma follows.

If, on the other hand, $x \geq 1/2$, then lower bound given for
the first statement of the lemma is not good since it may even
become negative. In this case we observe that the term $(1-x)^j
x^{k-j}$ is minimized for $j=k$. Setting $j=k$ in~(\ref{pol}) we
obtain the second statement of the lemma. \hfill$\Box$

\begin{corollary}
The following bounds hold for the time instance $t'$:

\noindent If $hx(0) < 1/2$ then
\begin{eqnarray}
t' \leq \frac{x(0) (h-1)}{l [hx(0))^k + 1 - 2hx(0)]}.
\label{tsmall}
\end{eqnarray}

\noindent If $hx(0) \geq 1/2$ then
\begin{eqnarray}
t' \leq \frac{k x(0) (h-1)}{l[1-hx(0)]^k {k \choose \lceil
\frac{k}{2} \rceil} \lceil \frac{k}{2} \rceil}. \label{tlarge}
\end{eqnarray}

\end{corollary}
\noindent From~(\ref{tsmall}) we see that the percentage of
similarities grows fast, if we start from $x(0)$ aiming at
$hx(0)$, with $h \geq 1$ and $ hx(0) < 1/2$ (and~(\ref{tsmall}) is
only a very coarse upper bound). If the target is, however, at
$hx(0)$ with $x \geq 1/2$ the upper bound is not good as the
denominator tends to 0 fast. However, since this denominator is
simply the first derivative of $x(t)$ at $hx(0)$, this derivative
fast tends to 0 if $hx(0) \geq 1/2$ and, thus, the tangent at this
point tends to become parallel to the $t$-axis. Thus, we have
again fast convergence.

\section{Conclusions}
\label{concl}

In this paper we described a series of protocols that can be used
in order to increase the percentage of similarities between two
strings held by two communicating parties without revealing their
values. The protocols involve the examination of random position
subsets of size $k$ and the flipping of a randomly chosen subset
of these positions if the two strings are found to disagree in
more than half the positions. In this way, the random process
governing the protocol is directed towards flipping disagreement
positions more than the flipping of agreement positions. The
propose protocols are, in fact, general and may be used in any
situation involving either wireless or conventional networks in
which there is no trusted third party or key management authority
among the network nodes.

\newpage

\begin{center}
\noindent {\Large APPENDIX}

\noindent {\bf Approximating stochastic processes with
deterministic functions: Wormald's Theorem~\cite{Wor99}}
\end{center}

\noindent
\begin{definition}
A function $f$ satisfies a Lipschitz condition on $D \subset
\Re^j$ if there exists some constant $L > 0$ such that $$
|f(u_1,\ldots,u_j) - f(v_1,\ldots,v_j)| \leq L \sum_{i = 1}^j |u_i
- v_i| $$ for all $(u_1,\ldots,u_j)$ and $(v_1,\ldots,v_j)$ in
$D$.
\end{definition}

\begin{definition}
Given a random variable $X$ depending on $n$, denoted by
$X^{(n)}$, we say that $X^{(n)} = o(f(n))$ always if
$$\max \{x | \prob[X^{(n)} = x] \neq 0\} = o(f(n)).$$
\end{definition}

\begin{theorem}
Let $Y_i^{(n)}(t)$, $n \geq 1$, be a sequence of real-valued
random variables, $1 \leq i \leq k$ for some fixed $k$, such that
for all $i$, all $t$ and all $n$, $|Y_i^{(n)}(t)| \leq Bn$ ($n >
0$) for some constant $B$. Let ${\bf H}(t)$ be the history of the
sequence, i.e. the matrix $\langle
\overrightarrow{Y}(0),\ldots,\overrightarrow{Y}(t) \rangle$, where
$\overrightarrow{Y}(t) = (Y_1^{(n)}(t),\ldots,Y_k^{(n)}(t))$.

Let $I = \{(y_1,\ldots,y_k): \prob[\overrightarrow{Y}(0)
=(y_1n,\ldots,y_kn)] \neq 0 \mbox{ for some } n\}$. Let $D$ be
some bounded connected open set containing the intersection of
$\{(s,y_1,\ldots,y_k): s \geq 0\}$ with a neighborhood of
$\{(t/n,y_1,\ldots,y_k): (y_1,\ldots,y_k) \in I\}$. (That is,
after taking a ball around the set $I$, $D$ is required to contain
the part of the ball in the half-space corresponding to $s = t/n$,
$s \geq 0$.)

Let $f_i: \Re ^{k+1} \rightarrow \Re, 1 \leq i \leq k$, and
suppose that for some $m = m(n)$,

\begin{itemize}

\item[(i)]
for all $i$ and uniformly over all $t < m$, always
\begin{eqnarray*} & & \ex[Y_i^{(n)}(t+1) - Y_i^{(n)}(t) | {\bf H}(t)] \\ & & = f_i(t/n,
Y_0^{(n)}(t)/n,\ldots, Y_k^{(n)}(t)/n) + o(1),\end{eqnarray*}
\item[(ii)]
for all $i$ and uniformly over all $t < m$,

$$ \prob[|Y_i^{(n)}(t+1) - Y_i^{(n)}(t)| > n^{1/5}] = o(n^{-3}),
\mbox{ always}, $$
%
\item[(iii)]
for each $i$, the function $f_i$ is continuous and satisfies a
Lipschitz condition on $D$.
\end{itemize}

Then

\begin{itemize}
\item[(a)]
for $(0,\hat{z}^{(0)},\ldots,\hat{z}^{(k)}) \in D$ the system of
differential equations
$$ \frac{dz_i}{ds} = f_i(s,z_0,\ldots,z_k), 1 \leq i \leq k $$
has a unique solution in $D$ for $z_i: \Re \rightarrow \Re$
passing through $z_i(0) = \hat{z}^{(i)}, 1 \leq i \leq k$, and
which extends to points arbitrarily close to the boundary of $D$;

\item[(b)]
almost surely $ Y_i^{(n)}(t) = z_i(t/n) \cdot n + o(n),$
uniformly for $0 \leq t \leq \min\{\sigma,m\}$ and for each $i$,
where $z_i(s)$ is the solution in~(a) with $\hat{z}^{(i)} =
Y_i^{(n)}(0)/n$, and $\sigma = \sigma(n)$ is the supremum of those
$s$ to which the solution can be extended.
\end{itemize}
%
\label{wormald}
\end{theorem}

\end{document}